\documentclass[11pt]{elsart}
\usepackage[dvips]{graphicx}

\usepackage{array}
\usepackage{color}
\usepackage{amssymb}
\usepackage{amsmath}
\usepackage{hhline}
\usepackage{longtable}
\usepackage{dcolumn}
\usepackage{bm}
\usepackage{subfigure}

\begin{document}
\begin{frontmatter}

\title{Deterministic Weighted Scale-free Small-world Networks}

\author[ad1]{Yichao Zhang}
\author[ad2,ad3]{Zhongzhi Zhang}
\ead{zhangzz@fudan.edu.cn}
\author[ad2,ad3]{Shuigeng Zhou}
\ead{sgzhou@fudan.edu.cn}
\author[ad1]{Jihong Guan\corauthref{G}}
\corauth[G]{Corresponding author.}
\ead{jhguan@mail.tongji.edu.cn}
\address[ad1]{Department of Computer
Science and Technology, Tongji University,
\\4800 Cao'an Road, Shanghai 201804, China}%
\address[ad2]{Department of Computer Science and Engineering, Fudan
University,
\\Shanghai 200433, China}%
\address[ad3]{Shanghai Key Lab of Intelligent Information Processing, Fudan
University,
\\Shanghai 200433, China}%

\begin{abstract}
We propose a deterministic weighted scale-free small-world model for
considering pseudofractal web with the coevolution of topology and
weight. Considering the fluctuations in traffic flow constitute a
main reason for congestion of packet delivery and poor performance
of communication networks, we suggest a recursive algorithm to
generate the network, which restricts the traffic fluctuations on it
effectively during the evolutionary process. we provide a relatively
complete view of topological structure and weight dynamics
characteristics of the networks: weight and strength distribution;
degree correlations; average clustering coefficient and
degree-cluster correlations; as well as the diameter.

\end{abstract}


\date{}
\end{frontmatter}


\section{Introduction}

To understand the general principles in architectures of networks,
many deterministic models are introduced into complex
networks~\cite{PA299559,PRE71036144,PRE65066122,PRE69037104,PRE65056101,sci2971551,PRE67026112,PRE67045103,PRE71036132,PRL94018702,PA363567,PA367613,NJP726,PRE73066126,EPL8610006,PRE78052103,PRE71046141,PRE74046105,PRE71056131,PRE056207,PA3561}.
These models are useful tools for investigating analytically not
only topological features of networks in
detail~\cite{PA299559,PRE71036144,PRE65066122,PRE69037104,PRE65056101,sci2971551,PRE67026112,PRE67045103,PRE71036132,PRL94018702,PA363567,PA367613},
but also dynamical problems on the
networks~\cite{NJP726,PRE73066126,EPL8610006,PRE78052103}. Before
presenting our own findings, it is worth reviewing some of this
preceding work to understand its achievements and shortcomings.
Deterministic scale-free networks were firstly proposed by
Barab\'asi \emph{et al.} in Ref.~\cite{PA299559} and intensively
studied in Ref.~\cite{PRE71036144} to generate a scale-free
topology. However, to some extent, the small exponent $\gamma$ of
the degree distribution for the model didn't satisfy the real
statistic results well. Instead, Dorogovtsev~\emph{et al.}
introduced another elegant model, called pseudofractal scale-free
web (PSW)~\cite{PRE65066122} which is extended by Comellas \emph{et
al.} consequently~\cite{PRE69037104}. Based on a similar idea of
PSW, Jung \emph{et al.} presented a class of recursive
trees~\cite{PRE65056101}, which have the small-world behavior built
in. Additionally, in order to discuss modularity, Ravasz \emph{et
al.} proposed a hierarchical network
model~\cite{sci2971551,PRE67026112}, the exact scaling properties
and extensive study of which were reported in
Refs.~\cite{PRE67045103} and~\cite{PRE71036132}, respectively.
Recently, motivated by the problem of Apollonian space-filing
packing, Apollonian networks~\cite{PRL94018702} with a typical loop
structure were introduced and intensively investigated
\cite{PRE71046141,PRE74046105,PRE71056131,PRE056207,PA3561}. These
pioneering works are all invaluable tools for the topology of
networks studies.

In the last few years, it is found that many real networks are
inhomogeneous, consisting of distinct nodes and links. For instance,
the scientist collaboration network, where scientists are identified
with nodes, and an edge exists between two scientists if they have
coauthored at least one paper~\cite{PNAS1013747}, and the Internet
at the AS level, where the link weights represent the bandwidth of a
cable and node weight the load of a router~\cite{PRE65066130}, among
other areas. Recently, weight dynamics ideas have been applied with
success to topics as diverse, such as random walks~\cite{NJP9154},
condensation~\cite{PRE77066105}, synchronization~\cite{PRL96164102},
traffic congestion~\cite{PRE79026112}, epidemic
spreading~\cite{PRE73036116,cpl22510}, information
filtering~\cite{PRE76046115}, to name but a few. The findings above
might provide insight for understanding the correlations among
weighted quantities and the underlying topological structure and
dynamics behaviors of the weighted networks.

Most previous weighted random
models~\cite{PRE70066149,PRL92228701,PRE75026111,PRL94188702,PRE71066124}
with topology and weight coevolution, however, possess very loose
clustering structures when the size of the networks is large. At the
same time, previous deterministic models, are mainly
unweighted~\cite{PA299559,PRE71036144,PRE65066122,PRE69037104,PRE65056101,sci2971551,PRE67026112,PRE67045103,PRE71036132,PRL94018702,PA363567,PA367613,NJP726,PRE73066126,EPL8610006,PRE78052103,PRE71046141,PRE74046105,PRE71056131,PRE056207,PA3561},
which ignore the heterogeneity of edges in real networks. What's
more, the models~\cite{PRE65066122,PRE69037104} on PSW networks fail
to provide the reason for adopting the recursive way to build up the
networks. Consequently, in this paper, we introduce a model bringing
weight evolution into the growth of pseudofractal scale-free web
(PSW)~\cite{PRE65066122} that aims to circumvent these incongruities
properly. As we will show, in the case of the recursive
construction, the traffic and its fluctuation decrease exponentially
with time either on edges or at nodes. Hence, we believe the
construction method may shed some light on networks design to
improve the control and speed of the whole
network~\cite{IEEETrans21}. At the same time, our comprehensive and
rigorous solutions may help people understand better the interplay
between network topology and weight dynamics.

\section{The model}

\begin{figure}
\begin{center}
\scalebox{0.8}[0.8]{\includegraphics{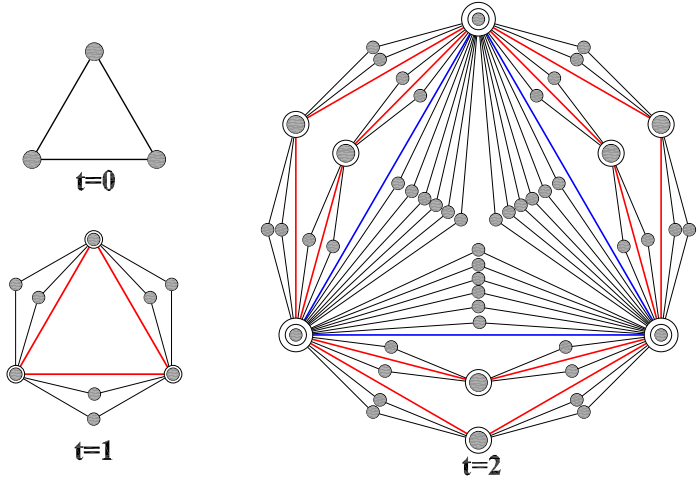}} \caption{Illustration
of the deterministically growing network for the particular case of
$m=2$ and $\delta=1$, showing the first three steps of growing
process. The gray links in the figure denote the links with weight
1, the red links with weight 3, and the blue links with weight 9.
The number of rings around a gray node denotes its age.}
\label{net01}
\end{center}
\end{figure}

The construction of the model is controlled by two parameters $m$
and $\delta$, evolving in a recursive way. We denote the network
after $t$ steps by $G(t)$, $t\geq 0$ (see Fig.~\ref{net01}). Then
the network at step $t$ is constructed as follows. For $t=0$, $R(0)$
is a triangle consisting of three links with unit weight. For $t\geq
1$, $G(t)$ is obtained from $G(t-1)$. We add $mw$ ($m$ is positive
integer) new nodes for each of the links with weight $w$, and we
connect each new node to both ends of this link by new links of unit
weight; moreover, we increase the weight of these links by $m\delta
w$ ($\delta$ is positive integer).

Before introducing our model further, we explain why adopting such a
recursive way and why the generated network is increasingly
efficient for transmitting information with network order. In this
model, the recursive construction is motivated by the practical need
to improve the transport capacity of real networks. As is known to
us, both the physical networks and the numbers of users are growing
continuously. The performance of the networks for larger system
sizes and heavier loads are critical issues to be addressed in order
to guarantee networks' functioning in he near future. For example,
if the traffic fluctuates dramatically, a highway is more likely to
be congested frequently when the peak value of traffic exceeding its
capacity. There is thus a need to build up more branches to
distribute the heavy traffic. However, ``how many branches must we
have?" and ``Where shall we put them?" are open questions yet.

Recently, the authors of Ref.~\cite{PRL100208701} claimed the
fluctuations in traffic flow constitute the main factor affecting
the performance of networks. They derived the dependence of
fluctuations with the mean traffic on unweighted networks
analytically. Consequently, their recipes were adopted extensively
to the weighted networks by the authors of
Ref~\cite{arXiv:0902.3160v3}. As shown in
Ref~\cite{arXiv:0902.3160v3}, for correlated networks (assortative
or disassortative mixing~\cite{prl89208701,pre67026126}), the
average traffic through a link $L_{ij}$ during a time window can be
represented as
\begin{equation}
\langle
f_{ij}\rangle=\frac{2\,w_{ij}}{\sum_{i=1}^Ns_i}RM,\label{fijcorrelated}
\end{equation}
and the standard deviation can be expressed as
\begin{equation}
{\sigma^2_{ij}}=\langle f_{ij}\rangle\left(1+\langle
f_{ij}\rangle\frac{\Delta^2+\Delta}{3R^2}\right),\label{edgesigma}
\end{equation}
where $w_{ij}$ is its link weight~\cite{arXiv:0902.3160v3}. The
length of the time window for observation is $M$. The average number
of cars or walkers among various time windows in the network is
denoted by $R$. $\Delta$ is defined as a random variable
representing the number of walkers travelling through the link in
the time window.

With respect to the traffic at nodes, the average traffic at a node
$i$
\begin{equation}
\langle
f_i\rangle=\frac{s_i}{\sum_{i=1}^Ns_i}RM,\label{ficorrelated}
\end{equation}
Then the standard deviation as a function of $\langle f_i\rangle$
can be obtained as:
\begin{equation}
{\sigma^2_i}=\langle f_i\rangle \left(1+\langle
f_i\rangle\frac{\Delta^2+\Delta}{3R^2}\right).\label{nodesigma}
\end{equation}
At each time step, the traffic will be distributed dispersedly to
the newly built edges (nodes). The larger, the size of the
considered network is, the lower the fluctuations on each edge (at
each node) should be. Details of the analysis are provided in
section~\ref{Weight distribution} and~\ref{Strength distribution}.

Notice that there are in fact three limiting cases of the present
model. In the special case $m=1$ and $\delta=0$, it is reduced to
the pseudofractal scale-free web described in~\cite{PRE65066122}.
When $\delta=0$, it is a particular case of the geometric growth
networks discussed in~\cite{PA377329}. When $m=1$, it is the same as
the deterministic weighted networks proposed in~\cite{aip77629}.
Thus, vary parameters $m$ and $\delta$, we can study many crossovers
between these limiting cases.

Let us consider the total number of nodes $N_t$, the total number of
links $E_t$ and the total weight of all links $W_t$ in $G(t)$. The
number of nodes created at step $t$ is denoted by $n_v(t)$. Note
that the addition of each new node leads to two new links, so the
number of links generated at step $t$ is $n_e(t)=2\,n_v(t)$. By
construction, for $t\geq 1$, we have
\begin{equation}\label{nv01}
n_v(t)=mW_{t-1},
\end{equation}
\begin{equation}\label{Et01}
E_t=E_{t-1}+2n_v(t),
\end{equation}
and
\begin{equation}\label{Wt01}
W_t=W_{t-1}(1+m\delta)+2mW_{t-1}.
\end{equation}
On the right-hand side of Eq.~(\ref{Wt01}), the first item is the
sum of weight of the old links, and the second term describes the
total weight of the new links generated in step $t$.
Eq.~(\ref{Wt01}) can be simplified to
\begin{equation}\label{Wt02}
W_t=(1+m\delta+2m)W_{t-1}.
\end{equation}
Considering the initial condition $W_0=3$, we obtain
\begin{equation}\label{Wt03}
W_t=3\,(1+m\delta+2m)^{t}.
\end{equation}
Substituting Eq.~(\ref{Wt03}) into Eq.~(\ref{nv01}) and using
$W_0=3$, the number of nodes created at step $t$ ($t\geq 1$) is
obtained to be
\begin{equation}\label{nv02}
n_v(t)=3m(1+m\delta+2m)^{t-1}.
\end{equation}
Hence, one can figure out the growth of the network is accelerated.
Then the total number of nodes present at step $t$ is
\begin{equation}\label{Nt01}
N_t=\sum_{t_i=0}^{t}n_v(t_i)=\frac{3}{2+\delta}\,\left[(1+m\delta+2m)^{t}+\delta+1\right].
\end{equation}
Combining Eq.~(\ref{nv02}) with Eq.~(\ref{Et01}) and considering
$E_0=3$, it follows that
\begin{equation}\label{Et02}
E_t=\frac{3}{2+\delta}\,\left[(1+m\delta+2m)^{t}+\frac{\delta}{2}
\right].
\end{equation}
Thus for large $t$, the average degree $\overline{k}_t=
\frac{2E_t}{N_t}$ is approximately equal to $4$.

\section{Structural properties}
In what follows we will study how the tunable parameters $m$ and
$\delta$ control some relevant characteristics of the weighted
network $G(t)$. Firstly, we give out the analytical solution of
distribution of strength, degree and weight to test its scale-free
nature; simultaneously, we show the analytical expression of average
traffic and its deviation of nodes and edges; subsequently, we move
forward to the average clustering coefficient coupled with the
diameter of this network for the sake of verifing its small-world
property; finally, we study the degree correlations as well.

\subsection{Weight distribution}\label{Weight distribution}
Let $w_e(t)$ be the weight of link $e$ at step $t$. In the view of
all the links emerging simultaneously have the same weight, it can
be recast recursivel as follows
\begin{equation}\label{we01}
w_e(t)=(1+m\delta)w_e(t-1).
\end{equation}
If link $e$ enters the network at step $\tau$, then $w_e(\tau)=1$.
Thus, we can easily have
\begin{equation}\label{we02}
w_e(t)=(1+m\delta)^{t-\tau}.
\end{equation}
Obviously, the weight spectrum of the network is discrete. It
follows that the weight distribution is given by
\begin{equation}
P(w)=\left\{\begin{array}{lc} {\displaystyle{n_e(0)\over
E_t}={2+\delta\over 2(1+m\delta+2m)^{t}+\frac{\delta}{2}} }
& \ \hbox{for}\ \tau=0,\\
{\displaystyle{n_e(\tau)\over
E_t}={2m(2+\delta)\,(1+m\delta+2m)^{\tau-1}\over
 2(1+m\delta+2m)^{t}+\frac{\delta}{2}} }
& \ \hbox{for}\  \tau\ge 1,\\
0 & \ \hbox{otherwise}\end{array} \right.
\end{equation}
and that the cumulative weight distribution
\cite{siamr45167,PRE65066122} is
\begin{equation}
P_{\rm cum}(w)=\sum_{\mu \leq \tau}\frac{n_e(\mu)}{E_t}
={2(1+m\delta+2m)^{\tau}+\frac{\delta}{2} \over
2(1+m\delta+2m)^{t}+\frac{\delta}{2}}.
\end{equation}
Substituting for $\tau$ in this expression using $\tau=t-\frac{\ln
w}{\ln (1+m\delta)}$ gives
\begin{eqnarray}\label{gammaw}
P_{\rm cum}(w)&=&{2(1+m\delta+2m)^{t} w^{-\frac{\ln(1+m\delta+2m)}{\ln(1+m\delta)}}+\frac{\delta}{2} \over 2(1+m\delta+2m)^{t}+\frac{\delta}{2}}\nonumber\\
          &\approx&w^{-\frac{\ln(1+m\delta+2m)}{\ln(1+m\delta)}}\qquad \qquad \qquad \hbox{for large $t$}.
\end{eqnarray}
Apparently, the weight distribution follows a power law with the
exponent $\gamma_{w}=1+{\frac{\ln(1+m\delta+2m)}{\ln(1+m\delta)}}$.
For the particular case of $m=1$, Eq.~(\ref{gammaw}) recovers the
result previously obtained in Ref.~\cite{aip77629}.

In this paper, the networks generated by the model are
disassortative, which will be discussed analytically in
section~\ref{Degree correlations}. For disassortative networks, the
fluctuation $\sigma_{ij}$ on an edge $L_{ij}$ relies on average
traffic $f_{ij}$, while $f_{ij}$ is governed by
$\frac{2w_{ij}}{\sum_{i=1}^Ns_i}$ with $R$ and $M$ fixed. Inserting
Eq.~(\ref{we02}) and Eq.~(\ref{Wt03}) into
Eq.~(\ref{fijcorrelated}), one can obtain the average traffic on an
arbitrary edge $L_{ij}$ can be written as
\begin{equation}\label{fe}
\langle
f_{ij}\rangle=\frac{2\,w_{ij}(t)}{2\,W_t}RM=\frac{(1+m\delta)^{t-\tau_{ij}}}{3\,(1+m\delta+2m)^{t}}RM,
\end{equation}
where $\tau_{ij}$ denotes the entry time of $L_{ij}$. Thus the
standard deviation can be expressed as
\begin{eqnarray}\label{De}
{\sigma^2_{ij}}&=&\left(\frac{1+m\delta}{1+m\delta+2m}\right)^t\frac{(1+m\delta)^{-\tau_{ij}}RM}{3}\nonumber\\
&+&\left(\frac{1+m\delta}{1+m\delta+2m}\right)^{2t}\frac{(1+m\delta)^{-2\tau_{ij}}(\Delta^2+\Delta)M^2}{27},
\end{eqnarray}
Eq.~(\ref{fe}) and Eq.~(\ref{De}) show that both $f_{ij}$ and
$\sigma_{ij}$ decrease as an exponential function of time $t$ with
$R$ and $W$ fixed as a result of $\frac{1+m\delta}{1+m\delta+2m}<1$.
{\color{blue}Notice that, we consider the parameter $R$ as a const
with the time evolution in our model in that the relation between
$R(t)$ and $t$ depends on various specific systems, which is not the
focus of the present paper. In fact, one can easily observe that the
recursive algorithm can also restrict the traffic fluctuations on
edges in the considered networks effectively during the evolutionary
process, in which $R(t)$ is not larger than
$(\frac{1+m\delta+2m}{1+m\delta})^t$.} Discussion of traffic and its
fluctuation at nodes will subsequently be given in the next section.

\subsection{Strength distribution}\label{Strength distribution}
In a weighted network, a node strength is a natural generalization
of its connectivity. The strength $s_{i}$ of node $i$ is defined as
\begin{equation}\label{si01}
s_i = \sum_{j\in \Omega_{i}} w_{ij} \, ,
\end{equation}
where $w_{ij}$ denotes the weight of the link between nodes $i$ and
$j$, $\Omega_{i}$ is the set of all the nearest neighbors of $i$.
The strength distribution $P(s)$ measures the probability that a
randomly selected node has exactly strength $s$.

Let $s_i(t)$ be the strength of node $i$ at step $t$. If node $i$ is
added to the network at step $t_i$, then $s_i(t_i)=2$. Moreover, we
introduce the quantity $\Delta s_i(t)$, which is defined as the
difference between $s_i(t)$ and $s_i(t-1)$. By construction, one can
easily obtain
\begin{eqnarray}\label{Deltasi01}
\Delta s_i(t)&=&s_i(t)-s_i(t-1)=m\delta\sum_{j\in \Omega_{i}}
w_{ij}+m\sum_{j\in \Omega_{i}} w_{ij}\nonumber\\
&=& m\delta s_i(t-1)+ m\,s_i(t-1).
\end{eqnarray}
Here the first item accounts for the increase of weight of the old
links existing in step $t-1$. The second term describes the total
weigh of the new links with unit weight that are generated in step
$t$ and connected to $i$.

From Eq.~(\ref{Deltasi01}), one can derive following recursive
relation:
\begin{equation}\label{si02}
s_i(t)=(1+m\delta+m)s_i(t-1).
\end{equation}
Using $s_i(t_i)=2$, we obtain
\begin{equation}\label{si03}
s_i(t)=2\,(1+m\delta+m)^{t-t_i}.
\end{equation}
Since the strength of each node has been obtained explicitly in
Eq.~(\ref{si03}), we can get the strength distribution via its
cumulative distribution~\cite{PRE65066122,siamr45167}, i.e.
\begin{equation}\label{pcums01}
P_{\rm cum}(s)=\sum_{\mu \leq t_i}\frac{n_v(\mu)}{N_t}
={(1+m\delta+2m)^{t_i}+\delta+1\over (1+m\delta+2m)^{t}+\delta+1}.
\end{equation}
From Eq.~(\ref{si03}), we can derive $t_i=t-\frac{\ln (s/2)}{\ln
(1+m\delta+m)}$. Substituting the obtained result of $t_i$ into
Eq.~(\ref{pcums01}), we have
\begin{eqnarray}\label{gammas}
P_{\rm cum}(s)&=&{(1+m\delta+2m)^{t} \left(\frac{s}{2}\right)^{-\frac{\ln(1+m\delta+2m)}{\ln(1+m\delta+m)}}+\delta+1 \over (1+m\delta+2m)^{t}+\delta+1}\nonumber\\
          &\approx&\left(\frac{s}{2}\right)^{-\frac{\ln(1+m\delta+2m)}{\ln(1+m\delta+m)}}\qquad \qquad \qquad \hbox{for large $t$}.
\end{eqnarray}
Thus, node strength distribution exhibits a power law behavior with
the exponent
$\gamma_{s}=1+{\frac{\ln(1+m\delta+2m)}{\ln(1+m\delta+m)}}$. For the
special case $m=1$, Eq.~(\ref{gammas}) recovers the results
previously reported in Ref. \cite{aip77629}.

On the other hand, the fluctuation at an arbitrary node $i$ is based
on $\frac{s_i}{\sum_{i=1}^Ns_i}$. Inserting Eq.~(\ref{si03}) and
Eq.~(\ref{Wt03}) into Eq.~(\ref{fijcorrelated}), one can obtain the
average traffic at an arbitrary node $i$ can be represented as
\begin{equation}\label{fnode}
\langle
f_i\rangle=\frac{s_i}{2\,W_t}RM=\frac{(1+m\delta+m)^{t-t_i}}{3\,(1+m\delta+2m)^{t}}RM,
\end{equation}
where $t_i$ denotes the entry time of the node $i$. Then the
standard deviation as a function of $\langle f_i\rangle$ is
\begin{eqnarray}\label{sigmanode}
{\sigma^2_{i}}&=&\left(\frac{1+m\delta+m}{1+m\delta+2m}\right)^t\frac{(1+m\delta+m)^{-t_{i}}RM}{3}\nonumber\\
&+&\left(\frac{1+m\delta+m}{1+m\delta+2m}\right)^{2t}\frac{(1+m\delta+m)^{-2t_{i}}(\Delta^2+\Delta)M^2}{27}.
\end{eqnarray}
It is easy to find that both $f_{i}$ and $\sigma_{i}$ decrease
exponentially with $t$ as $\frac{1+m\delta+m}{1+m\delta+2m}<1$,
which is similar with the former result on edges.
{\color{blue}Although the strength of nodes growing exponentially,
the traffic and fluctuation at them still decrease with the growing
size of the networks in the case that $R$ is a constant or
$R(t)\leq(\frac{1+m\delta+2m}{1+m\delta})^t$. In other words, the
sufficient condition of keeping the potential traffic fluctuation
problems away from the resulting networks is
$R(t)\leq(\frac{1+m\delta+2m}{1+m\delta})^\frac{ln(\frac{N_t(2+\delta)}{3}-\delta-1)}{ln(1+m\delta+2m)}$
or the average number of walkers is invariable. The novel property
is interesting and has not been investigated by previous
works~\cite{PRE65066122,PRE70066149,PRL92228701,PRE75026111,PRL94188702,PRE71066124}.
Therefore, to some extent, this model may provide a paradigm to
control the traffic fluctuations and improve transport efficiency of
the whole network~\cite{IEEETrans21}.}

\subsection{Degree distribution}
\begin{figure}
\begin{center}
\includegraphics[width=9cm]{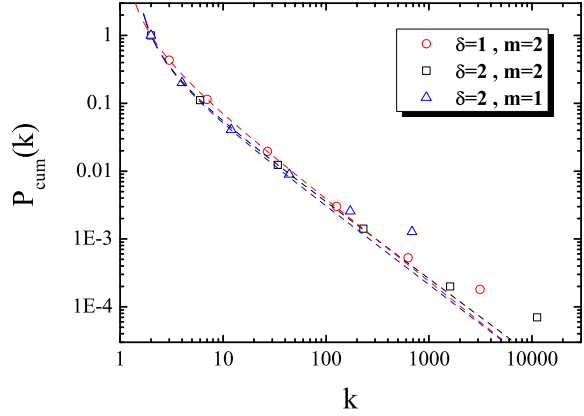}
\caption{Cumulative degree distribution $P_{cum}(k)$ versus $k$ for
different $\delta$ and $m$ corresponding to Eq.~(\ref{gammak}). The
measurements are taken at $t=5$, illustrating these networks display
a power-law degree distribution. The dashed lines are the best fits,
with $\gamma_{cum}=1.20806,-1.12887,-1.15699$ respectively.}
\label{PK}
\end{center}
\end{figure}

Similarly to the strength, all simultaneously emerging nodes have
the same degree.  Let $k_i(t)$ be the degree of node $i$ at step
$t$. If node $i$ is added to the network at step $t_i$, then by
construction $k_i(t_i)=2$. After that, the degree $k_i(t)$ evolves
as
\begin{equation}\label{ki01}
k_i(t)=k_i(t-1)+m\,s_{i}(t-1),
\end{equation}
where $ms_{i}(t-1)$ is the degree increment $\Delta k_i(t)$ of node
$i$ at step $t$. Substituting Eq.~(\ref{si03}) into
Eq.~(\ref{ki01}), we have
\begin{equation}\label{deltaki01}
\Delta k_i(t)=2m\,(1+m\delta+m)^{t-1-t_i}.
\end{equation}
Consequently, the degree $ k_i(t)$ of node $i$ at time $t$ is
\begin{equation}\label{ki02}
k_i(t)=k_i(t_i)+\sum_{\eta=t_i+1}^{t}{\Delta k_i(\eta)}
=2+\frac{2}{\delta+1}\left[(m\delta+1+m)^{t-t_i}-1\right].
\end{equation}
Analogously to computation of cumulative strength distribution, one
can find the cumulative degree distribution
\begin{eqnarray}\label{gammak}
P_{\rm cum}(k)
&={(1+m\delta+2m)^{t}\, \left[\frac{k}{2}\,(\delta+1)-\delta\right]^{-\frac{\ln(1+m\delta+2m)}{\ln(1+m\delta+m)}}+\delta+1\over (1+m\delta+2m)^{t}+\delta+1}\nonumber\\
&\approx
\left[\frac{k}{2}\,(\delta+1)-\delta\right]^{-\frac{\ln(1+m\delta+2m)}{\ln(1+m\delta+m)}}
\qquad \hbox{for large $t$}.
\end{eqnarray}
As is shown in the Fig.~\ref{PK}, the degree distribution has the
scale-free property with the same exponent as $\gamma_{s}$
($\gamma_{k}=\gamma_{s}=1+\gamma_{cum}$, where
$\gamma_{cum}={\frac{\ln(1+m\delta+2m)}{\ln(1+m\delta+m)}}$).

\subsection{Clustering coefficient}
In this model, the analytical expression for clustering coefficient
$C(k)$ of the individual node with degree $k$ can be derived
exactly. For instance, if a new node is connected to both ends of a
link,  its degree and clustering coefficient will be $2$ and $1$,
respectively. Naturally, its degree will increase by one when
connecting a new node in the next step. On the other hand, there
must be an existing neighbor of it attaching to the new node at the
same time. On the other hand, there must be an existing neighbor of
it, attaching to the new node as well. Because our networks are
corresponding to the particular case $q=2$ of the recursive clique
trees~\cite{PRE65066122,PRE69037104}, for a node of degree $k$, we
have
\begin{equation}\label{Ck}
C(k)= {{1+ (k-2)} \over {k(k-1)\over 2}}= 2/k.
\end{equation}
The scaling $C(k)\sim k^{-1}$ has been found for some network
models~\cite{PRE67026112,PRL94018702,PA367613,PRE71046141,PA3881713},
and has also observed in several real-life
networks~\cite{PRE67026112}.

Using Eq.~(\ref{Ck}), we can obtain the clustering $\overline{C}_t$
of the networks at step $t$:
\begin{equation}\label{AC}
\overline{C}_t=
    \frac{1}{N_{t}}\sum_{r=0}^{t}
    \frac{2n_v(r)}{k_r},
\end{equation}
where the sum is the total of clustering coefficient for all nodes
and $k_r=2+\frac{2}{(\delta+1)}\left[(m\delta+1+m)^{t-r}-1\right]$
shown by Eq.~(\ref{ki02}) is the degree of the nodes created at step
$r$.

\begin{figure}
\begin{center}
\includegraphics[width=9cm]{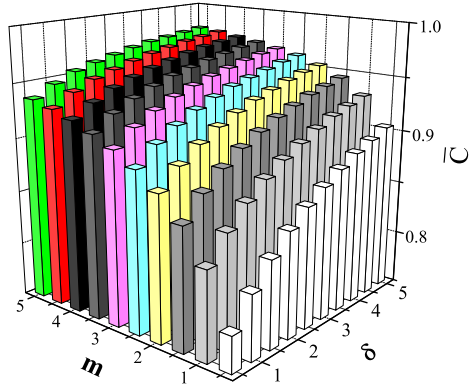}
\caption{The solutions of Eq.~(\ref{AC}) for $\delta$ and $m$
ranging from $1$ to $5$ respectively. The measurements are taken at
$t=100$, illustrating these networks display a high degree of
clustering.} \label{cc}
\end{center}
\end{figure}

It can be easily proved that $\overline{C}_t$ increases with $q$ for
arbitrary fixed $m$, and likely $\overline{C}_t$ increases with $m$
when $q$ fixed. In the case of $t=100$ ($N\rightarrow\infty$),
Eq.~(\ref{AC}) converges to a correspondingly large value
$\overline{C}$. When $\delta=2$, for $m=1,2,3,4$ and $5$,
$\overline{C}$ equal to $0.886,0.922,0.941,0.952$ and $0.96$.
respectively. When $m=2$, for $\delta=1,2,3,4$ and $5$,
$\overline{C}$ are $0.899,0.922,0.937, 0.947$ and $0.954$,
respectively. Evidently, the clustering coefficient of our networks
is correspondingly stable and close to $1$. Moreover, the average
clustering coefficient $\overline{C}$ can be tuned by $\delta$ and
$m$ (see Fig.~\ref{cc}).
\begin{figure}
\begin{center}
\scalebox{0.7}[0.7]{\includegraphics{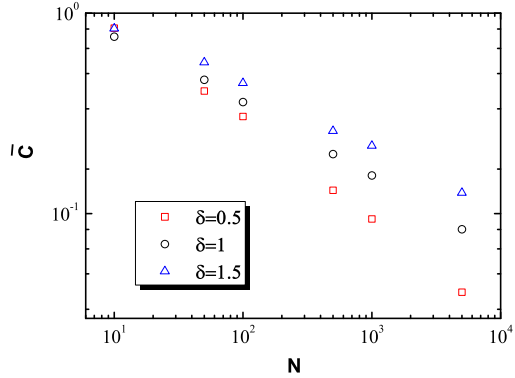}} \caption{Average
clustering coefficient $\overline{C}$ as a function of size $N$ of
nodes for different $\delta$ in the BBV networks are shown in the
inset, using the weight of a new link $w_0=1$, the size of initial
seed $m_0=3$, the degree of a new node $m=3$.}\label{BBV}
\end{center}
\end{figure}
\begin{figure}
\begin{center}
\scalebox{0.5}[0.7]{\includegraphics{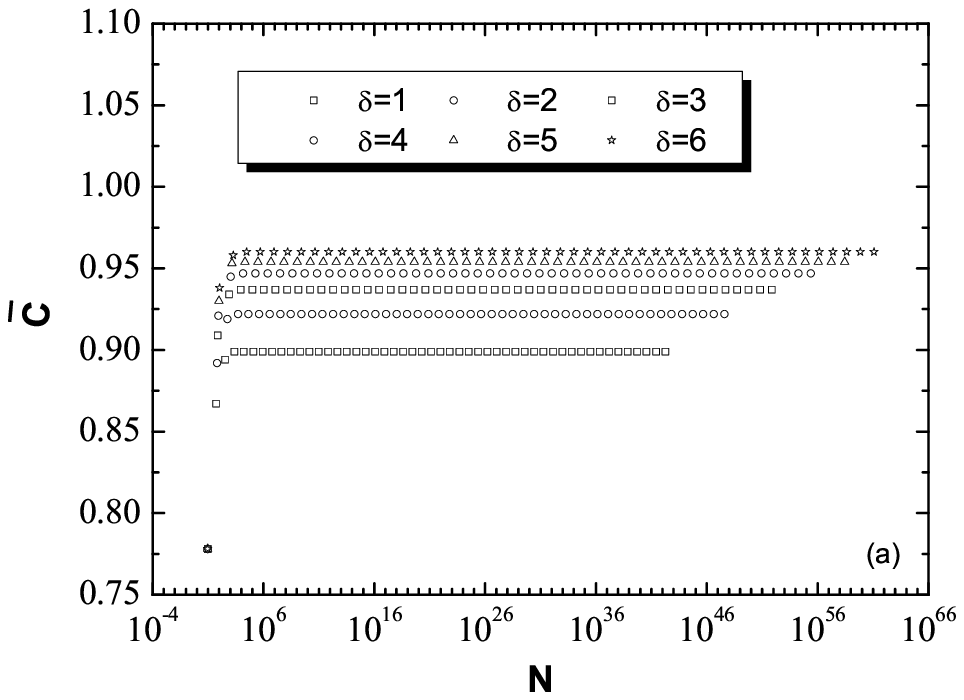}}
\scalebox{0.5}[0.7]{\includegraphics{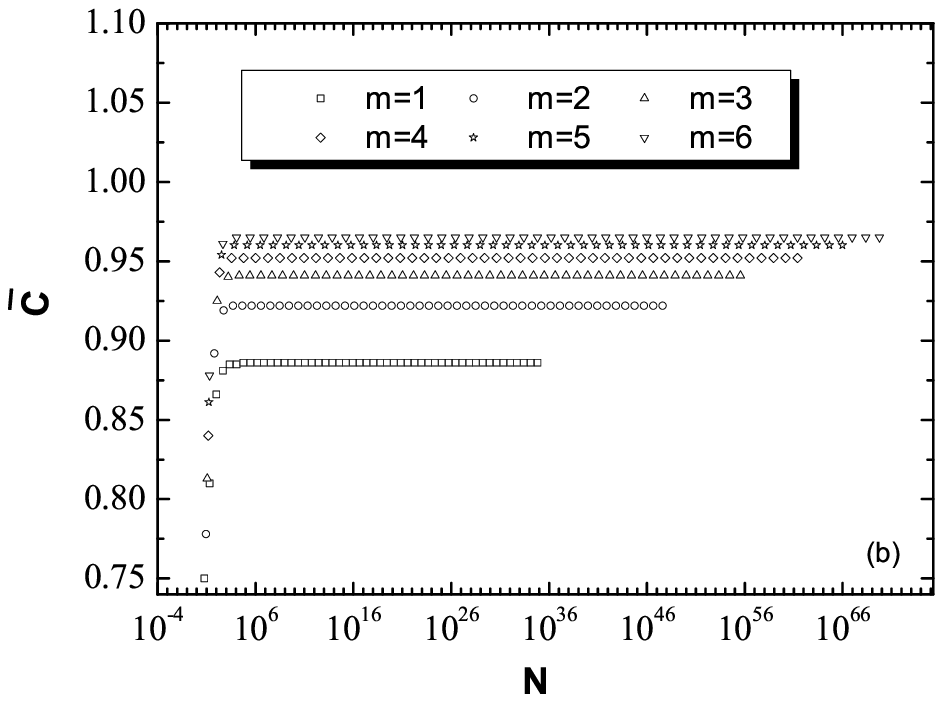}} \caption{Semilogarithmic
plots of the average clustering coefficient $\overline{C}$ against
networks size $N$ for (a) $\delta=1,2,...,6$ when $m=2$ and (b)
$m=1,2,...,6$ when $\delta=2$ corresponding to Eq.~(\ref{AC}). The
measurements are taken over a time interval $t=1,2,...,50$,
illustrating that $\overline{C}$ is a finite nonzero value
independent of $N$ for our model.} \label{FSS}
\end{center}
\end{figure}

In the classical weighted co-evolutionary models, for example, BBV
networks \cite{PNAS1013747,PRL92228701}, the average clustering
coefficient rapidly decreases when the networks is growing (see
Fig.~\ref{BBV}). Our simulations confirm that in the limit of large
networks ($N\gg1$), the BBV networks' clustering coefficient is
getting close to zero. However, many real-world networks have a
relatively stable and nonzero clustering coefficient, which make the
results of BBV model useful, but far from comprehensive. In the
Fig.~\ref{FSS}, we performed numerical solutions for our model at
various values of $\delta$ with fixed $m$ in the panel (a) (various
values of $m$ with fixed $\delta$ in the panel (b)) up to $t=50$.
For the infinite network, one can easily obtain that this tunable
average clustering coefficients of our networks is independent of
their size and tends to a nonzero limit, which is a unique property
shared by many real networks~\cite{sci2971551,siamr45167}. At the
same time, this feature gives excellent agreement with the previous
unweighted studies~\cite{PRE65066122}.

\subsection{Diameter}

As is known to all, the diameter of a network is defined as the
maximum of the shortest distances between all pairs of nodes, which
characterizes the longest communication delay in the network. Small
diameter is an important property of small-world network.
Fortunately, for our networks, it can be calculated easily. Below we
give the precise analytical computation of diameter of $G(t)$
denoted by $Diam(G(t))$.

It is easy to see that at step $t=0$ (resp. $t=1$), the diameter is
equal to $1$ (resp. $2$). At each step $t\geq2$, one can easily see
that the diameter always lies on a pair of nodes created at step
$t$. In order to simplify the analysis, we first note that it is
unnecessary to check all the nodes in the networks to fix the
diameter. In other words, some nodes (``inner" nodes) added at a
given step can be ignored, because they do not increase the diameter
of the previous net. Here, so-called ``inner" nodes are those that
connect to links that already existed before step $t-1$. Indeed, for
these nodes we know that a similar construction has been done in
previous steps, so they have nothing to do with the calculation of
the diameter.

Let us call ``outer" nodes the nodes which are connected to a fresh
link. Clearly, at each step, the diameter depends on the distances
between outer nodes. At any step $t\geq2$, we note that an outer
node cannot be linked with two nodes created during the same step
$r\leq t-1$. Indeed, we know that from step $2$, no outer node is
connected to two nodes of the initial triangle $G(0)$. Thus, for any
step $t\geq2$, any outer node is connected with nodes that appeared
at pairwise different steps. Now consider two outer nodes created at
step $t\geq2$, say $v_t$ and $w_t$. Then $v_t$ is connected to two
nodes, and one of them must have been created before or during step
$t-2$.

We summarize the above arguments, and gather them into two cases:
(a) $t=2l$ is even. Then, if we make $l$ `jumps' from $v_t$ we reach
the initial triangle $G(0)$ in which we can reach any $w_t$ by using
a link of $G(0)$ and making $l$ jumps to $w_t$ in a similar way.
Thus $Diam(G(2l))\leq2l+1=t+1$. (b) $t=2l+1$ is odd. In this case we
can stop after $l$ jumps at $G(1)$, for which we know that the
diameter is $2$, and make $l$ jumps in a similar way to reach $w_t$.
Thus $Diam(G(2l+1))\leq2(l+1)=t+1$. Obviously, the bound can be
reached by pairs of outer nodes created at step $t$. More precisely,
these two nodes $v_t$ and $w_t$ share the property, that both of
them are connected to two nodes added at steps $t-1$, $t-2$
respectively. Hence, formally, $Diam(G(t))=t+1$ for any $t\geq0$.
Considering $N_t\sim(1+m\delta+2m)^t$, the diameter is small and
scales logarithmically with the number of network nodes.

\subsection{Degree correlations}\label{Degree correlations}
In complex network, degree
correlations~\cite{prl89208701,pre67026126,sci296910,prl87258701,prl65066130},
has attracted much attention, because it can give out a unique
description of network structures, which could help researchers
understand the characteristics of
networks~\cite{prl89208701,prl87258701,prl65066130,NP2275}. An
interesting quantity related to degree correlations is the average
degree of the nearest neighbors for nodes with degree $k$, denoted
as $k_{nn}(k)$~\cite{prl87258701,prl65066130,prl89208701}. When
$k_{nn}(k)$ increases with $k$, it means that nodes have a tendency
to connect to nodes with a similar or larger degree. In this case
the network is defined as
assortative~\cite{prl89208701,pre67026126}. In contrast, if
$k_{nn}(k)$ is decreasing with $k$, which implies that nodes of
large degree are likely to have near neighbors with small degree,
then the network is said to be disassortative. If correlations are
absent, $k_{nn}(k)=const$.

In our networks, we can acquire $k_{nn}(k)$ exactly by
Eq.~(\ref{ki02}). Except for three initial nodes generated at step
$0$, no nodes born in the same step, will be linked to each other.
All links from the newcomers to old nodes with larger degree are
made at their creation steps. Then, these newcomers become old ones
to accept the nodes with smaller degree made at each subsequent
steps. These results are shown in the expression
\begin{eqnarray}\label{knn1}
k_{nn}(k) =\frac{1}{n_v(t_i)k(t_i,t)}
(\sum^{t'_i=t_i-1}_{t'_i=0}m\cdot
n_v(t'_i)s(t'_i,t_i-1)k(t'_i,t)\nonumber\\+\sum^{t'_i=t}_{t'_i=t_i+1}m\cdot
n_v(t_i)s(t_i,t'_i-1)k(t'_i,t)).
\end{eqnarray}
Here the first sum on the right-hand side accounts for the links
made to nodes with larger degree (i.e., $t'_i<t_i$) when the node
was generated at $t_i$. The second sum describes the links made to
the current smallest degree nodes at each step $t'_i>t_i$.

Substituting Eqs.~(\ref{nv02}) and~(\ref{ki02}) into
Eq.~(\ref{knn1}), one expects that
\begin{eqnarray}\label{knn2}
k_{nn}(t_i,t)\approx{2}[\frac{(m\delta+1+2m)(m\delta+1+m)}{m\delta^2+2m\delta+m+\delta}\nonumber\\
\cdot\frac{(m\delta+1+m)^{2t_i}}{(m\delta+1+2m)^{t_i}}+\frac{t-t_i}{m\delta+1+m}],
\end{eqnarray}
in the infinite limit of $t$, where
$k_r\approx\frac{2}{\delta+1}(m\delta+1+m)^{t-t_i}$. In another
word, the initial step $k_{nn}(t_i,t)$ grows linearly with time.
Consequently, writing Eq.~(\ref{knn2}) in terms of $k$, it is
straightforward to obtain
\begin{eqnarray}
k_{nn}(k,t)\approx{2}\frac{(m\delta+1+2m)(m\delta+1+m)}{m\delta^2+2m\delta+m+\delta}\frac{(m\delta+1+m)^{2t}}{(m\delta+1+2m)^t}
\nonumber\\\cdot
\left[\frac{k-2}{2}\,(\delta+1)+1\right]^{-\frac{2\ln(1+m\delta+m)-\ln(1+m\delta+m)}{\ln(1+m\delta+m)}}.
\end{eqnarray} Apparently, $k_{nn}(k)$ is approximately a power law
function of $k$ with negative exponent, which indicates that the
networks are disassortative. Note that $k_{nn}(k)$ of the Internet
exhibits a similar power-law dependence on the degree $k_{nn}(k)\sim
k^{-w}$, with $w=0.5$~\cite{prl87258701}.

\section{Conclusion and discussion}
To sum up, we have proposed and investigated a deterministic
weighted network model, which is constructed in a recursive fashion.
The recursive construction guarantees that the traffic fluctuations
of nodes and edges decrease exponentially with the time of
evolution. The weights of these networks characterizing the various
connections exhibit complex statistical features with highly tunable
degree, strength, and weight distributions, which display power-law
behavior. We have shown the analytical results for degree
distributions with tunable exponent and large clustering
coefficient, as well as small diameter. Particularly, the features
of clustering coefficient in our proposed model, i.e., it is
independent of its net size, might lead to a better understanding of
realistic networks. To some extent, our model can thus perform well
in controlling and designing a variety of weighted scale-free
small-world networks to improve their transport efficiency.
\smallskip

\section*{Acknowledgment}
This research was supported by the National Basic Research Program
of China under grant No. 2007CB310806, the National Natural Science
Foundation of China under Grant Nos. 60704044, 60873040 and
60873070, Shanghai Leading Academic Discipline Project No. B114, and
the Program for New Century Excellent Talents in University of China
(NCET-06-0376).


\end{document}